\documentclass[fleqn,10pt]{wlscirep}

\usepackage[utf8]{inputenc}
\usepackage[T1]{fontenc}
\usepackage{float}
\usepackage{subcaption}
\usepackage{tcolorbox}
\usepackage{placeins}
\usepackage{amsmath}
\usepackage{mathtools}
\usepackage{chngcntr}

\title{Scalable Drift Monitoring in Medical Imaging AI}

\author[1,*,$\dag$]{Jameson Merkow}
\author[2,3,$\dag$]{Felix J. Dorfner}
\author[2,3]{Xiyu Yang}
\author[1]{Alexander Ersoy}
\author[2,3]{Giridhar Dasegowda}
\author[2,3]{Mannudeep Kalra}
\author[1]{Matthew P. Lungren}
\author[2,3,\ddag,+]{Christopher P. Bridge}
\author[1,\ddag]{Ivan Tarapov}

\affil[1]{Microsoft Health and Life Sciences (HLS), Redmond, WA, USA}
\affil[2]{Department of Radiology, Massachusetts General Hospital, Boston, MA, USA}
\affil[3]{Harvard Medical School, Boston, MA, USA}
\affil[*]{jameson.merkow@microsoft.com}
\affil[+]{corresponding author: cbridge@mgh.harvard.edu}

\affil[$\dag$]{these authors contributed equally to this work}
\affil[$\ddag$]{co senior authors}

\newcommand{\metric}{\psi}

\newcommand{\metricos}{\metric}

\newcommand{\mmc}{\mathrm{MMC}}

\newcommand{\sshift}{\zeta}

\newcommand{\sscale}{\eta}

\newcommand{\sweight}{\alpha}

\newcommand{\detwindow}{W}
\newcommand{\refwindow}{R}

\newcommand{\stride}{\delta}
\newcommand{\duration}{\tau}

\newcommand{\distsamp}{d_{n}}
\newcommand{\distset}{D}
\newcommand{\distisamp}{d_{n}^{(c)}}

\begin{abstract}

The integration of artificial intelligence (AI) into clinical medical imaging has advanced diagnostics but also poses challenges in maintaining long-term reliability as input data change. To address this, we present MMC+, a modular framework for scalable drift monitoring that builds on our CheXstray framework for real-time drift detection using multi-modal data concordance (MMC). MMC+ provides a flexible architecture that enables cost-effective continuous performance monitoring in real-world healthcare settings. The framework's capabilities include aggregating information from multiple data streams using statistical tests or distance metrics, leveraging foundation model embeddings to remove the need for site-specific encoder training, and quantifying uncertainty to better communicate drift significance. Validated with data from Massachusetts General Hospital during the COVID-19 pandemic, MMC+ effectively detects significant data shifts that correlate with performance changes and indicates when AI systems may deviate from acceptable performance bounds.
The code for MMC+ is publicly available (Link: \url{https://github.com/FJDorfner/MedImaging-ModelDriftMonitoring/tree/MMC_plus}), promoting broader adoption of AI solutions in clinical settings. Our framework significantly advances drift detection while promoting transparency and collaboration in clinical AI research.

\end{abstract}
\begin{document}
 
\flushbottom
\maketitle

\thispagestyle{empty}

\section{Introduction}

The role of artificial intelligence (AI) in medical imaging has expanded significantly in recent years as evidenced by the increasing number of academic publications and the rising adoption of AI applications in clinical settings \cite{van2021artificial, west2019global, benjamens2020state, mehrizi2021applications, tadavarthi2020state}. To ensure the safe and effective use of these AI algorithms, researchers and clinicians are developing a broad set of best practices for evaluation and deployment. These practices include ``day 0'' validation, which typically involves evaluation using a curated retrospective dataset to assess AI model performance prior to deployment\cite{park2023methods, omoumi2021buy}. As the field matures and clinical AI governance practices are being developed, there is also growing recognition of the need for continuous, explainable monitoring tools that mitigate the risks of prediction drift, feature drift, and input data shifts after a model is deployed\cite{Feng2022,Rajpurkar2023,Pianykh2020}. Drift arises commonly in imaging due to factors including changes in equipment, clinical protocols, and patient demographics, and recent case studies have highlighted the risk that such drift can lead to significant degradation of model accuracy following deployment\cite{Lacson2022,Rahmani2023,ross2022ai}. 

Notably, the FDA has underscored the importance of robust, real-time monitoring in AI-enabled medical devices, mandating adaptive oversight to ensure these systems remain safe and effective as clinical environments evolve\cite{fda2023ai}. In its recent guidance on AI-enabled device software functions, the FDA further emphasizes the necessity for comprehensive post-market surveillance strategies, including continuous performance tracking, early detection of data and feature drifts, and dynamic recalibration protocols.

Despite the recognized importance of continuous and explainable monitoring, many organizations rely on direct user reporting of model performance \cite{daye2022implementation,bizzo2023addressing}. While this offers an effective means to assess whether the model's predictions remain accurate, it frequently fails to identify the root cause of performance changes such as changes in the input data. Moreover, direct performance monitoring requires timely access to ground truth labels, which are often provided manually, and as such monitoring continuously in this manner remains cost-prohibitive and may diminish the efficiency benefits of AI. If instead checks are performed periodically, the result may be delayed detection and response to problems. Furthermore, in some situations, including prognostic models, immediate access to ground truth is impossible and other methods of model monitoring are therefore a necessity.

A promising approach to overcoming the limitations of direct performance monitoring involves monitoring for the changes in the statistical properties of input data streams and model outputs using fully automated methods. Previous studies and methodologies have made significant strides in the monitoring of machine learning models in this way. However, these primarily focus on other domains, for example, the monitoring of Electronic Health Record (EHR) data \cite{subasri2022cyclops, duckworth2021using} and lack support for imaging data. Supporting imaging data often requires generating image embeddings to create low-dimensional summaries of images for the purpose of drift detection, with traditional approaches like Variational Autoencoders (VAEs) \cite{kingma2013auto} posing several challenges. VAEs require extensive site-specific training, lack the expressiveness to capture the complexity of medical images, and may necessitate separate models for different modalities or locations. This increases operational complexity and computational burden, limiting scalability and adaptability across institutions with diverse imaging protocols.

Our prior work with CheXstray \cite{merkow_chexstray_2023} addresses this gap by developing a framework to support the automated monitoring of medical imaging AI models designed to include comprehensive image data analysis. CheXstray introduced a novel metric called multi-modal concordance (MMC), which aggregates metrics from various sources to assess overall data drift, and crucially does not assume access to ongoing ground truth labels (beyond an initial reference window for calibration) to monitor model performance. In an experimental setting, MMC demonstrated a strong correlation with model performance when artificial drift was induced within a datastream. However, recent research has shown that real-world data is more complex, and relying solely on surrogate metrics like MMC for performance monitoring can be ineffective \cite{kore_empirical_2024}. 

Building on our previous conference paper \cite{merkow_chexstray_2023}, this work presents several key contributions to the field of AI model monitoring for medical imaging. We present MMC+, an extension that solidifies the framework and includes design principles for effective drift monitoring systems.
The contributions of this manuscript include (1) identifying essential properties that drift monitoring components should possess for robust deployment across diverse clinical settings; (2) demonstrating how foundation models provide image representations that satisfy scalability requirements; (3) conducting a real-world validation during the COVID-19 pandemic at a single academic medical center to evaluate these design principles; (4) analyzing the relationship between drift signals and model performance to understand when monitoring indicates the need for intervention; and (5) presenting an operational approach showing how characterized baseline ranges guide decisions about performance audits versus system stability. These insights provide guidance for architecting drift monitoring systems in medical imaging AI.

The implications of these findings are significant for the field of medical imaging AI, where data shifts can occur due to a variety of factors, such as changes in patient demographics, disease prevalence, and imaging protocols. By leveraging data drift monitoring, we can gain insight into different factors that drive change in model performance, helping us identify root causes and address gaps in many current monitoring solutions. This approach aligns with the need for a systematic, real-time performance monitoring framework that can adapt to the unique challenges posed by medical imaging data, ultimately supporting the broader adoption and integration of AI solutions in healthcare settings and simplifying regulatory compliance by providing traceable, auditable performance metrics that align with FDA requirements.

\section{Results}

\begin{figure}[ht]
    \centering
    \includegraphics[width=1\linewidth]{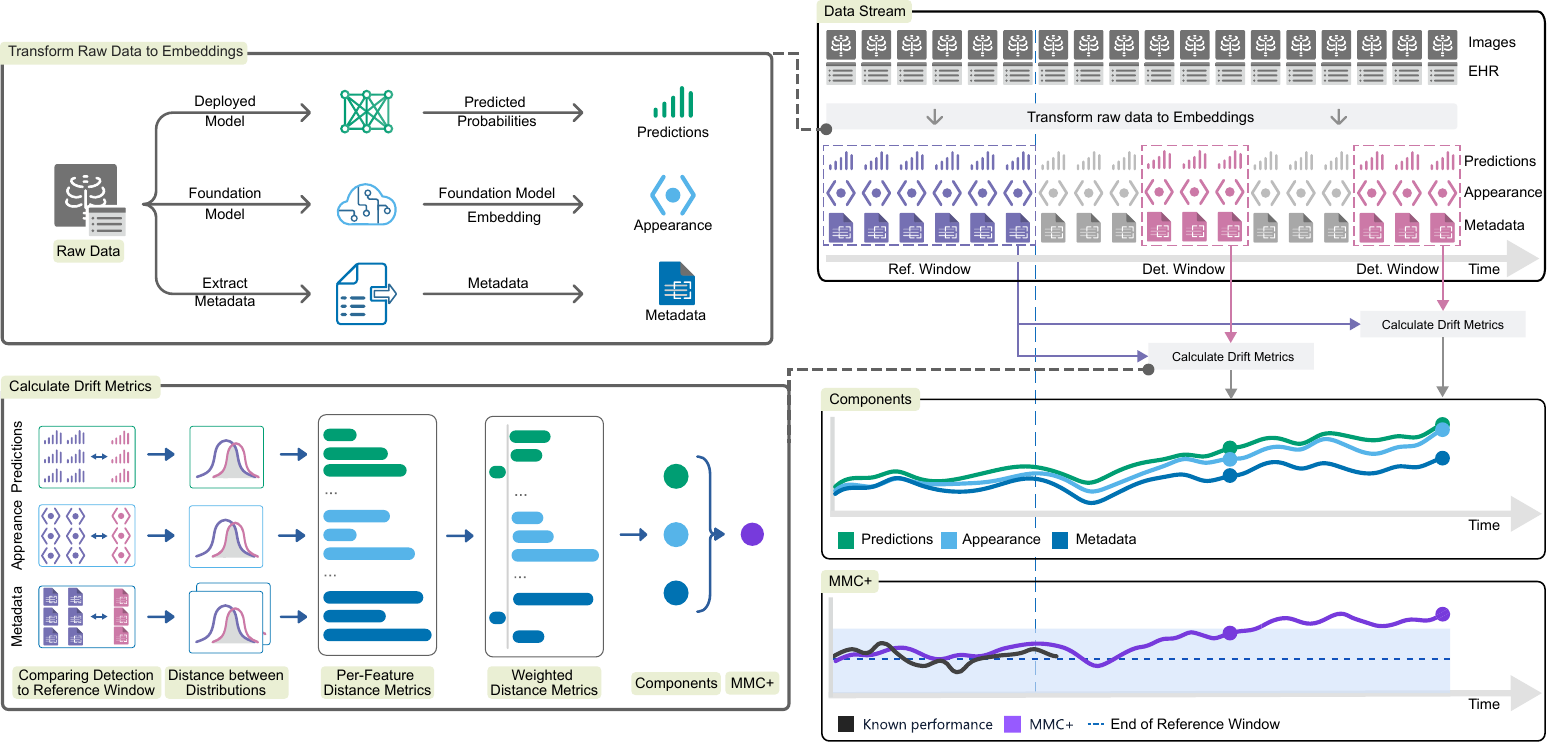}
    \caption{
    Overview of the MMC+ framework for drift monitoring.
    Right: the monitoring pipeline. A continuous stream of imaging studies and
    associated EHR data (\textit{Data Stream}) is transformed into three derived
    data streams: model predictions, image appearance, and metadata. A sliding
    detection window (pink) is compared against a pre-determined reference
    window (purple) on which model performance is well characterized. For each
    of the three components, a distance metric quantifies the distributional
    dissimilarity between the two windows; the resulting drift metrics are
    tracked over time (\textit{Components}) and aggregated into the unified MMC+
    metric (\textit{MMC+}), shown alongside model performance, which is known
    only where ground-truth labels are available. The shaded band indicates the
    range of MMC+ values under stable conditions, and the vertical dashed line
    marks the end of the reference window. High MMC+ values indicate that the
    model is operating on data on which its performance is not well
    characterized. Left: detailed views of two pipeline steps (dashed
    connectors). \textit{Transform Raw Data to Embeddings}: each study is
    processed by the deployed model to yield predicted probabilities, by a
    foundation model to yield an appearance embedding, and by metadata
    extraction; these three representations constitute the derived data streams
    on which drift is subsequently quantified. \textit{Calculate Drift Metrics}:
    for each component, per-feature distance metrics are computed between the
    distributions of these representations in the reference and detection
    windows, then normalized and weighted before aggregation into the three
    component drift scores and the single MMC+ value.
    }
    \label{fig:mmcplus}
\end{figure}

Our proposed MMC+ metric, shown in Figure~\ref{fig:mmcplus}, quantifies the dissimilarity between data in a given temporal window and a well-characterized pre-determined reference window, on which model accuracy is known (see the \nameref{sec:methods} section). The higher the value of MMC+ in a given temporal window, the more dissimilar the data in that window is from the reference set. MMC+ aggregates three sources of information to give a unified dissimilarity score: image metadata, image appearance (using embeddings from the MedImageInsight foundation model \cite{codella2024medimageinsight}), and model outputs. MMC+ refines our earlier MMC metric~\cite{merkow_chexstray_2023} in several practical ways. Where the original MMC compared distributions using statistical hypothesis tests, MMC+ instead uses distance metrics that quantify distributional dissimilarity directly. This separates the effect size of a shift from its statistical significance, giving a more stable and interpretable signal that does not saturate with large sample sizes and remains well-defined when previously unseen categorical values appear. MMC+ further reports an uncertainty range rather than a single point estimate, and derives image-appearance features from a pretrained foundation model rather than a site-specific autoencoder, removing the need to retrain the encoder for each new deployment. The methodological details of these changes are described in the \nameref{sec:mmc} section.

We conducted various experiments to demonstrate the use of MMC+ on a representative AI model operating on a large, real-world, consecutive dataset. The AI model is a DenseNet-121 convolutional neural network (CNN) trained to detect nine findings (atelectasis, cardiomegaly, edema, lung opacity, pleural effusion, pleural other, pneumonia, pneumothorax, and support devices) independently in an input radiograph following the CheXpert convention \cite{Irvin_2019_chexpert}, see the \nameref{sec:classifier} section for more details. 
The dataset consists of 90,581 chest radiograph studies from 60,062 patients, collected from Massachusetts General Hospital (MGH) between July 2019 and June 2021, representing all chest radiographs acquired during that time window at MGH. The data captures variability across clinical settings, including a mix of inpatient, emergency, and outpatient studies. We divided this dataset into training (July–September 2019), reference (October–December 2019), and test (January 2020–June 2021) sets to mirror a real-world model development and deployment process. Labels for the nine findings were generated from the accompanying radiology reports using a large language model. This dataset setup allowed us to evaluate MMC+ across different temporal and clinical shifts (see the \nameref{sec:dataset} section).

For all the experiments presented in the following sections, we used a window duration $\duration$ of 30 days and a window stride $\stride$ of 1 day (see the \nameref{sec:mmc} section). We chose this window duration as a balance between minimizing lag time and addressing the noise and other issues that arise with smaller sample sizes, even though this choice can introduce some delay in drift detection. The number of resampling iterations, $N_s$, was set to $20$ (see the \nameref{sec:mmc} section).

\subsection{MMC+ Results on Real-World Data}

To evaluate the effectiveness of the MMC+ framework in detecting data drift, we applied MMC+ to a real-world dataset of chest radiographs collected over an 18-month period that includes the onset of the COVID-19 pandemic. This experiment aims to simulate a realistic scenario where an AI model faces significant shifts in data distribution, allowing us to assess how MMC+ responds to such changes and whether it can serve as a reliable indicator for initiating performance audits or model updates.

\begin{figure}[ht]
    \centering
    \includegraphics[width=.9\textwidth]{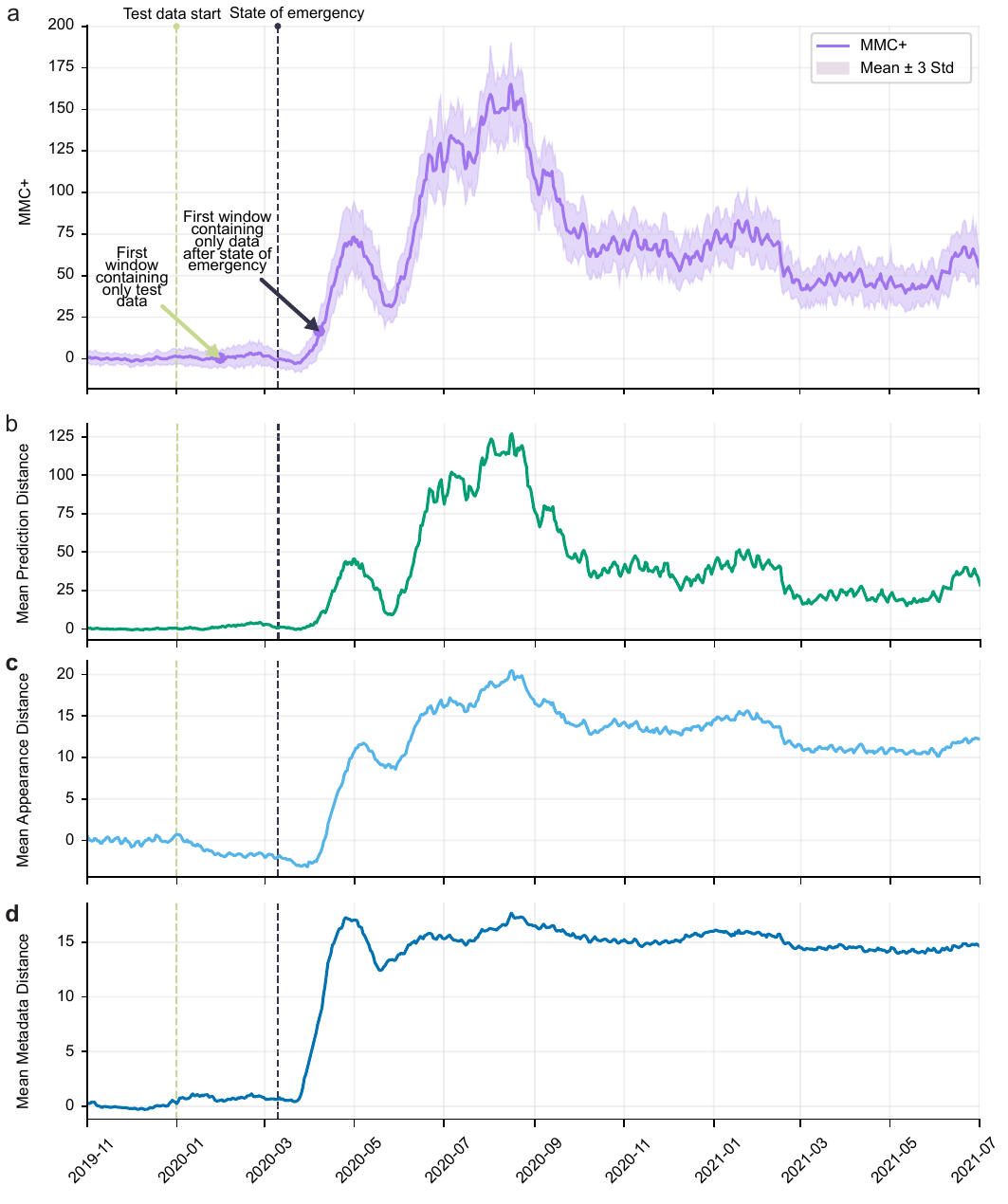}
    \caption{Results of the MMC+ experiment on chest radiographs from MGH. \textbf{a:} Plot of the evolution of MMC+ over time from 2019-11 to 2021-07. The purple line depicts the weighted MMC+ value and the purple shaded area indicates the range or variability around the MMC+, providing the uncertainty in the MMC+ measurement during this time period.  Two dashed vertical lines are shown: the first marks the start of the test data, and the second represents March 10, 2020, the day Massachusetts declared a state of emergency. Two arrows highlight the first windows where all data is sourced from either the test data or post-state-of-emergency period, a consequence of the window duration. \textbf{b-d:} The MMC+ metric is composed of three high level components that capture data drift: the model prediction metric (b) captures shifts in model predictions from a trained classifier providing changes in the model's output over time, the appearance metric (c) represents changes in the appearance of medical images as encoded by the MedImageInsight foundation model, and the metadata metric (d) measures variations in the metadata extracted from the DICOM files and RIS systems. All three components rise together in mid-March 2020 and remain elevated, indicating that the shift originates across the entire input data stream rather than from a single source.}
    \label{fig:mmc}
\end{figure}

Figure~\ref{fig:mmc} shows the MMC+ over time starting from November 1st, 2019, which marks the end of the first complete 30-day window within the reference set, until July 1st, 2021.
The MMC+ metric remains steady for approximately the first 2 months but increases rapidly in mid-March 2020. This also coincides with the declaration of the state of emergency in the state of Massachusetts on March 10, 2020 due to the COVID-19 pandemic (\url{https://www.mass.gov/info-details/covid-19-state-of-emergency}).
Thereafter, MMC+ fluctuates but always remains high, suggesting that the data never returns to its initial characteristics following the onset of the pandemic.
From  Figure~\ref{fig:mmc}  we see how the following individual component groups contribute to MMC+: metadata, image appearance, and model output, see the \nameref{sec:mmc-components} section for more details. All of these component groups show similar trends, suggesting all groups detect the dramatic changes.

\begin{figure}
    \centering
    \includegraphics[width=0.95\linewidth]{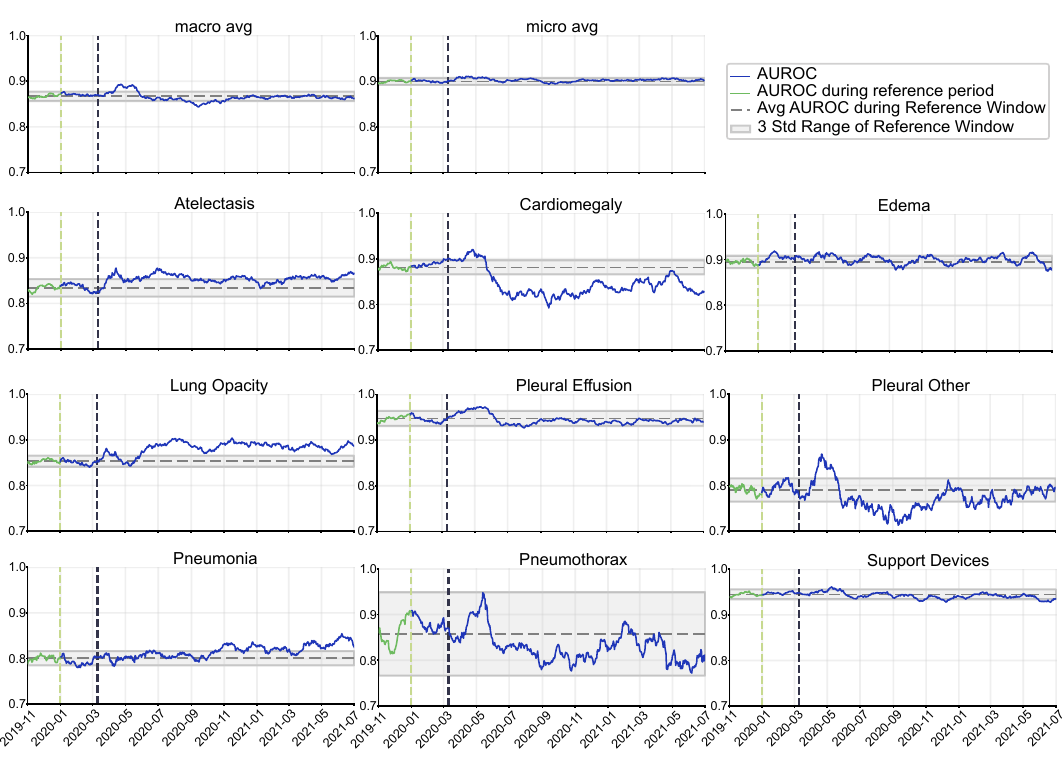}
    \caption{Model performance (AUROC) for the nine findings and the macro/micro averages. AUROC is marked in green within the reference period and blue after the reference period, with the average performance during the reference period shown as a solid grey line and the reference period's 3 standard deviation range as a grey shaded area for each finding.}
    \label{fig:performance}
\end{figure}

Next, we examine the model's performance in Figure~\ref{fig:performance}, which depicts the AUROC for the individual findings across the same time period. We use AUROC as a threshold-independent, prevalence-agnostic summary of classifier separability and as the reference performance signal against which MMC+ is assessed. We notice that nearly all performance indicators show some change starting in mid-March 2020, though they vary in scale and direction.

Starting with the most significant changes, we see that the performance for cardiomegaly drops significantly and remains low throughout the period. In contrast, ``pleural other'' initially shows an increase in performance but then dips around June 2020. We also observe that the performance for lung opacity detection improves starting in May 2020, despite significant changes in the data.

The remaining differences in performance are more subtle but still apparent. We see minor increases in performance for atelectasis and pneumonia. Findings such as edema, pleural effusion, support devices, and both averages (macro and micro) exhibit performance swings in both directions during this time period, indicating instability in model predictions.

Regarding pneumothorax performance, we observe large fluctuations within the performance curve. However, since the performance of predicting this finding also varied widely in the reference set, it is challenging to determine if these changes are connected to data drift or are due to inherent variability in the model's predictions for this condition.

These observations highlight the complex relationship between data drift and model performance. While MMC+ effectively detects shifts in the data distribution, the impact of this drift on model performance is not uniform across different findings.

\subsection{Operating Threshold and Performance Relationship}

A drift monitor is most useful when paired with an operating threshold that separates stable from drifted detection windows. This threshold is difficult to set empirically. The reference window's statistics are, by definition, the stationary baseline against which drift is measured, so the reference data cannot itself contain drift-positive examples to calibrate against. A threshold could be chosen retrospectively, once drift has occurred, but such a threshold has limited operational utility. It is therefore operationally useful to be able to set a threshold prospectively. In the \nameref{sec:mmc} section, we describe our approach to doing so and arrive at an operating threshold of 10, above which we consider a detection window to be under drift. We evaluate this prospective threshold along two complementary axes: 1) how closely it matches the threshold that would have been chosen optimally in retrospect, and 2) because drift matters operationally only insofar as it affects model performance, how cleanly it separates windows of stable from unstable performance.

To evaluate drift-event classification, we used the Massachusetts state-of-emergency declaration on March 10, 2020 as a proxy onset date. Because each detection window spans 30 days, windows immediately after this date contain a mixture of pre- and post-onset studies; we therefore repeated the calculation after shifting the positive-label boundary by 7, 15, and 30 days. We compared the prospective operating threshold of 10 with thresholds selected retrospectively after the drift labels were defined (Table~\ref{tab:mmc_fp_fn_counts}).

At \(\mmc+ = 10\), MMC+ produced no false-positive windows for the +0, +7, or +15 day onset definitions. The +30 day definition produced three such windows, all falling between March 10 and April 9, 2020, the state-of-emergency declaration and the delayed positive-label boundary for that onset definition. Because the state-of-emergency declaration is itself a proxy for the true onset of drift rather than a ground-truth boundary, drift-related shifts in the data stream likely predate the official notice; under this conservative delayed-onset definition, these windows could be interpreted as early detections rather than operational false alarms. False-negative counts decreased from 27/479 windows at +0 days to 0/449 windows at +30 days, consistent with the lag introduced by a 30-day monitoring window that initially contains both pre- and post-onset studies. Of the two retrospective thresholds, 11.7 is Youden-optimal for the +0, +7, and +15 day onset definitions and reproduces the same classification as the prospective threshold of 10 in every row; 19.3 is Youden-optimal specifically for the +30 day definition, where it achieves perfect separation (0 false positives, 0 false negatives), but at the cost of higher false-negative counts for the earlier onset definitions (27\(\to\)30, 20\(\to\)23, 12\(\to\)15). Taken together, the prospective threshold matches the post hoc optimum for the +0, +7, and +15 day definitions, before the transition period has fully resolved, and departs from it only at +30 days, where the additional flagged windows are consistent with the early-detection interpretation described above rather than a loss of specificity.

Next, we examined the relationship between MMC+ and performance consistency, assessing how MMC+ values correspond to the classifier's normalized AUROC across the nine findings, as well as their micro and macro averages. Performance variability was observed across all datastreams, which prompted the calculation of ``normal'' fluctuations based on AUROC values from the reference set (see the \nameref{sec:mmc} section for more details). These values were collected in 30-day windows to calculate the mean and standard deviation, which were subsequently used to normalize AUROC during the test period. 

\begin{figure}
    \centering
    \includegraphics[width=0.95\linewidth]{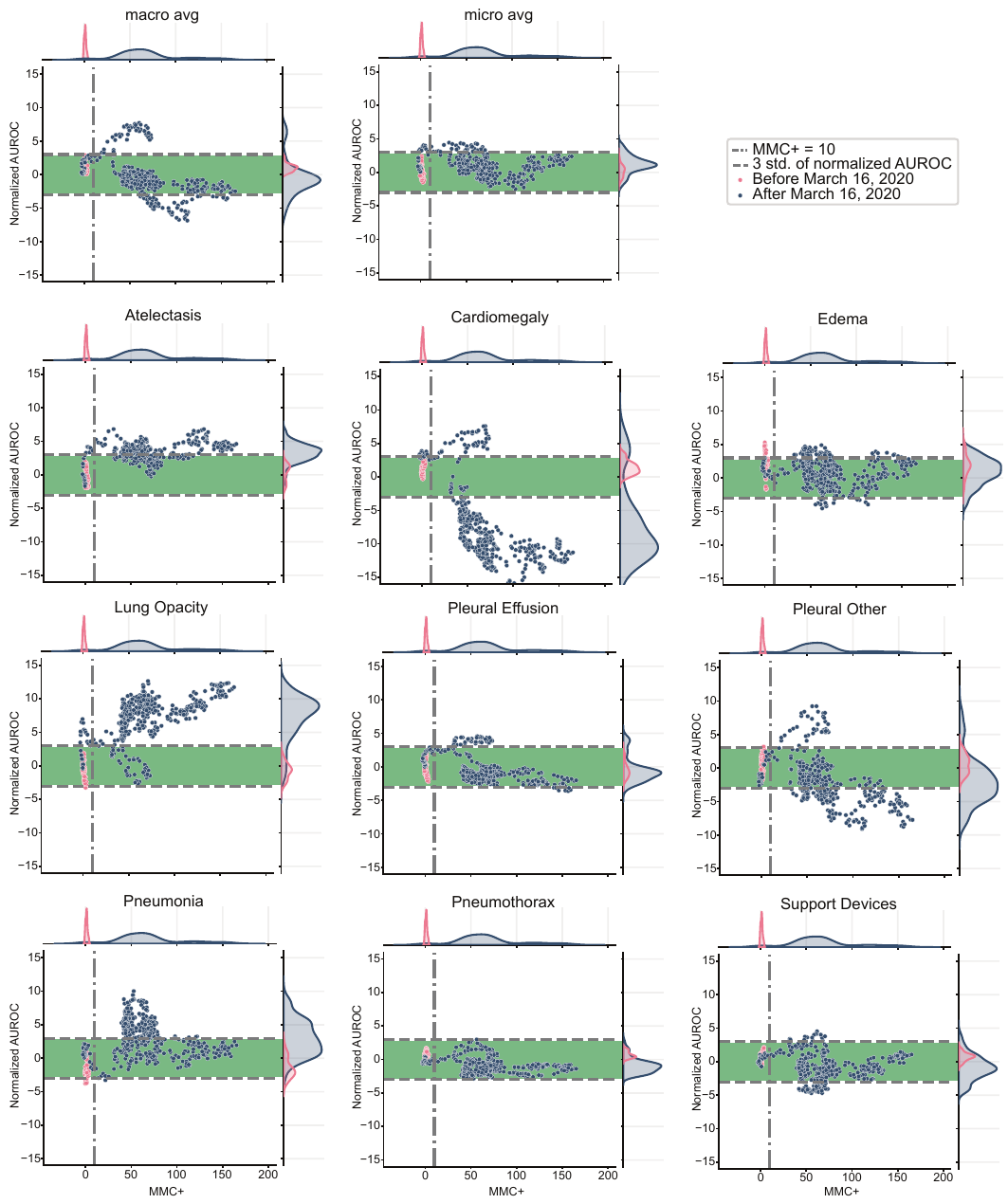}
    \caption{Relationship between MMC+ and normalized AUROC for each finding. AUROC is expressed in standard deviations from the reference-period mean. In each plot, the pink points represent test set data before March 10, 2020, while the blue points represent data after this date. The horizontal dotted lines mark $\pm3$ standard deviations of the AUROC, and the vertical line at $\mmc+ = 10$ represents the threshold where drift is observed. Above each scatter plot, KDE plots for the MMC+ and normalized AUROC are displayed, showing the distribution of test data pre- and post-March 10 (pink for pre, blue for post). These plots highlight the separation between the two time periods based on both MMC+ and AUROC behavior.}
    \label{fig:mmc_auroc_relationship}
\end{figure}

Each plot in Figure~\ref{fig:mmc_auroc_relationship} presents a scatter plot where the $x$-axis represents the MMC+ value and the $y$-axis corresponds to the normalized AUROC for a different label. Data points are color-coded based on whether they occur before or after the onset of the pandemic (defined as the March 10 declaration of emergency in the state of Massachusetts), with the edges of each scatter plot displaying a kernel density estimate (KDE) of the two groups to highlight any separation.

A clear pattern emerges from these figures: when MMC+ values remain below the prospective operating threshold of 10 (see the \nameref{sec:mmc} section), the normalized AUROC remains consistently within 3 standard deviations of the reference set mean. However, when MMC+ exceeds this threshold, the likelihood of the normalized AUROC remaining within 3 standard deviations of the reference mean diminishes significantly. Although performance does not universally decline (performance for some labels, such as atelectasis, even improves in certain instances), the variability in performance increases notably. This indicates that while MMC+ remains within its normal range, performance consistency is maintained, but outside this range, the model's reliability becomes less predictable. In other words, high MMC+ is a sensitive but not specific indicator of performance shifts.

These observations are further supported by summary statistics presented in Table~\ref{tab:mmc_auroc_stability}. In nearly all cases, we observe greater variation in performance when MMC+ is outside its normal range. Furthermore, in all cases except for Pneumothorax, which varies a great deal in the reference set, and Edema, where the rate changes only modestly (0.845 to 0.810), there is an increased proportion of AUROC values that deviate more than 3 standard deviations from the reference mean, underscoring the importance of maintaining MMC+ within its optimal range for consistent model performance.
\begin{table}[ht]
    \centering
    \small
    \setlength{\tabcolsep}{4pt}
    \renewcommand{\arraystretch}{0.98}
    \hspace*{-0.12\textwidth}%
    \begin{subtable}[t]{0.66\textwidth}
    \centering
    \caption{Drift-event classification}
    \label{tab:mmc_fp_fn_counts}
    \begin{tabular}{lcccc}
        \hline
        GT onset & {$n$ (pos / neg)} & FP/FN@10 & FP/FN@11.7 & FP/FN@19.3\\
        delay & & (prosp.) & (post.) & (post.)\\\hline
        +0d & 479 / 234 & 0 / 27 & 0 / 27 & 0 / 30 \\
        +7d & 472 / 241 & 0 / 20 & 0 / 20 & 0 / 23 \\
        +15d & 464 / 249 & 0 / 12 & 0 / 12 & 0 / 15 \\
        +30d & 449 / 264 & 3 / 0 & 3 / 0 & 0 / 0 \\
        \hline
    \end{tabular}
    \end{subtable}%
    \begin{subtable}[t]{0.31\textwidth}
    \centering
    \caption{AUROC stability}
    \label{tab:mmc_auroc_stability}
    \begin{tabular}{lcc}
        \hline
        Label &  MMC+ < 10 &  MMC+ >= 10\\\hline
        Atelectasis &  0.907 & 0.385 \\
        Cardiomegaly & 0.711 & 0.049 \\
        Edema & 0.845 & 0.810 \\
        Lung Opacity & 0.825 & 0.100 \\
        Pleural Other & 0.979 & 0.487 \\
        Pleural Effusion & 1.000 & 0.885 \\
        Pneumonia & 0.907 & 0.507 \\
        Pneumothorax & 1.000 & 1.000 \\
        Support Devices & 1.000 & 0.836 \\
        \hline
    \end{tabular}
    \end{subtable}
    \caption{Operating-threshold evaluation for MMC+. \textbf{(a)} False-positive (FP) and false-negative (FN) window counts for binary drift-event classification across four onset-delay assumptions. All rows are computed over 713 one-day-stride detection windows. The threshold of \(\mmc+ = 10\) is the prospective operating point; 11.7 and 19.3 are retrospective thresholds. \textbf{(b)} Rates at which AUROC remains within 3 standard deviations of the reference set AUROC for each finding below and above the prospective MMC+ operating threshold.}
    \label{tab:mmc_auroc_relationship}
\end{table}

\subsection{Utility of Metadata}

\begin{figure}[ht]
    \centering
    \includegraphics[width=.95\textwidth]{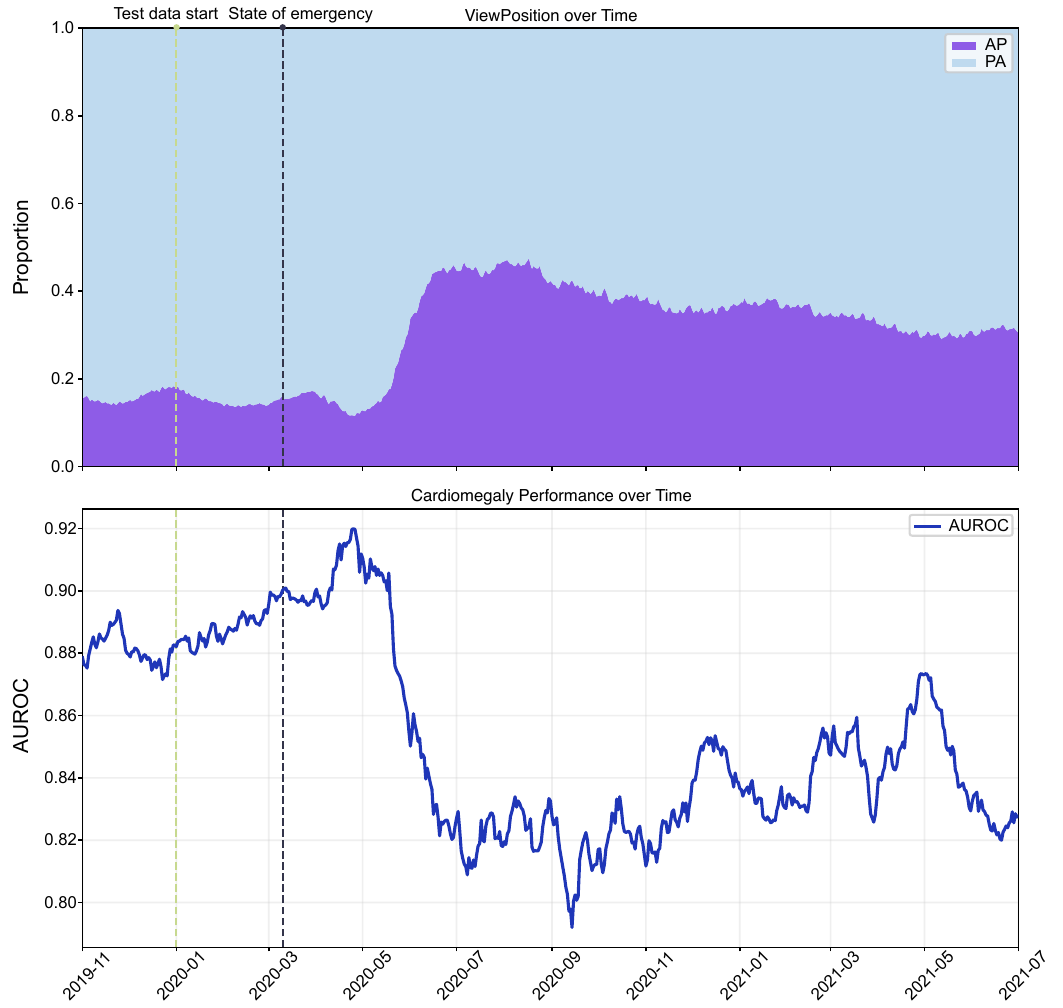}
    \caption{Radiograph view position and cardiomegaly detection performance. (top) View Position distribution (AP, purple, vs PA, blue) and cardiomegaly AUROC (bottom) over the same time period. The dashed vertical line marks March 10, 2020. This shared time axis shows simultaneous shifts in the view position distribution and cardiomegaly classification performance.}
    \label{fig:view_position}
\end{figure}
\begin{figure}[ht]
    \centering
    \includegraphics[width=.95\textwidth]{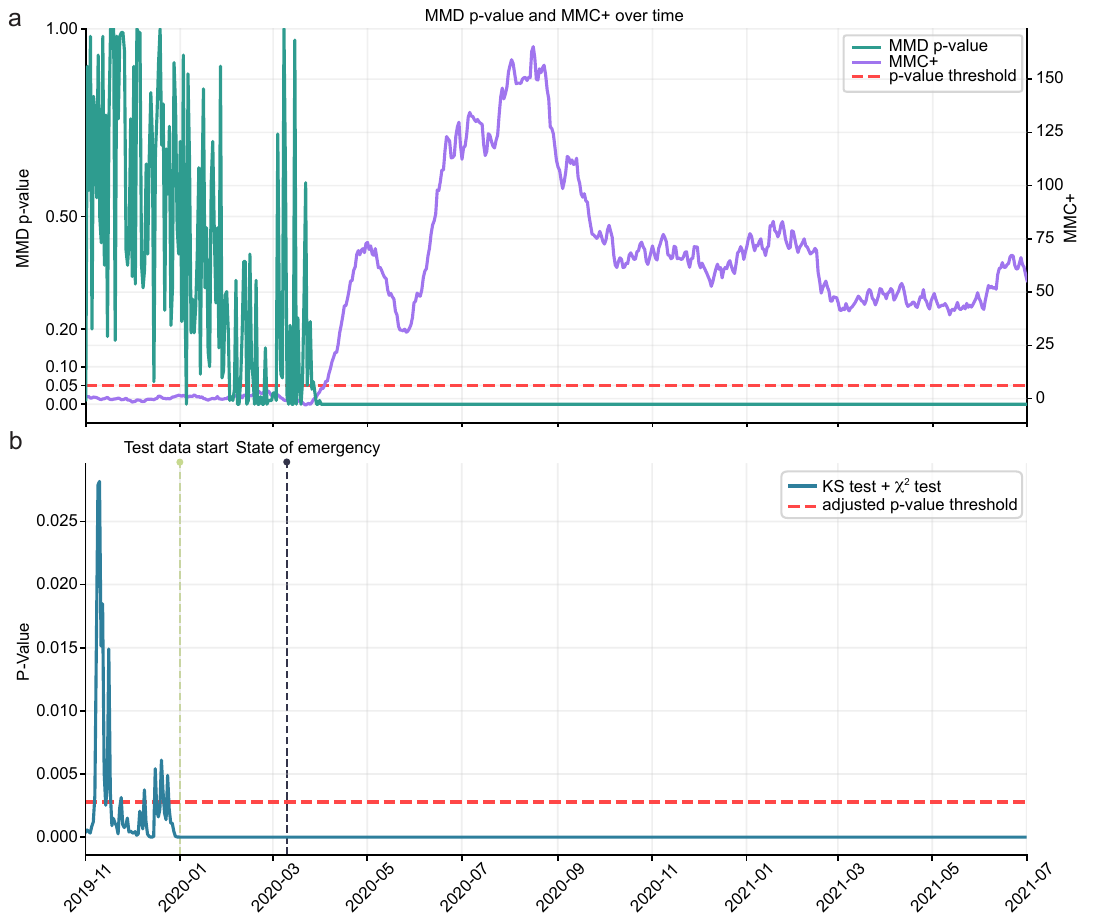}
    \caption{Comparison to $p$-value based metrics. a) MMD-based $p$-value (light blue) and MMC+ (purple) over the same period. Each time the $p$-value falls below $0.05$ (red dashed line) would trigger a drift notification. b) $p$-value calculated using KS tests (continuous variables) and $\chi^2$ tests (discrete variables) and combined using a Bonferroni correction. The red dashed line gives the corrected threshold below which drift notifications would be triggered. Due to conflation of effect size and significance, $p$-values are volatile and regularly cross the threshold even when no substantial shifts have occurred, producing frequent false alarms and leading to day-to-day ambiguity in whether a drift has actually occurred. By contrast, MMC+ provides a smooth signal that rises only on major shifts, allowing for reliable operational insights.}
    \label{fig:mmd_pavlue}
\end{figure}

Figure~\ref{fig:view_position} demonstrates the importance of incorporating metadata attributes sourced from both DICOM files and the Radiology Information System (RIS) into the monitoring framework (see the \nameref{sec:mmc-components} section). Specifically, the figure highlights a significant shift in the distribution of the ``View Position'' attribute for X-ray acquisitions starting around June 2020. Initially, less than 20\% of the images were captured using the antero-posterior (AP) projection, but by mid-2020, this proportion had increased to over 40\%. The heart and mediastinum appear larger on AP images than on PA images due to the increased distance to the image receptor and beam divergence, which negatively impacts the diagnostic quality of these images for cardiomegaly, an enlarged heart. Figure~\ref{fig:view_position} shows that this shift in the proportion of the view position is accompanied by a notable decline in the model’s performance for detecting cardiomegaly.

\subsection{Comparison to p-value Based Methods}
 
Several other methods for monitoring data drift track the $p$-value of a statistical hypothesis test between the data within a detection window and that in the reference set.
For example, Kore et al.\cite{kore_empirical_2024} monitor data drift using $p$-values from a maximum mean discrepancy (MMD) hypothesis test, and open-source drift detection toolkits Alibi-Detect \cite{alibi-detect} and Evidently AI \cite{evidently} both implement primarily $p$-value based metrics.
We compare to two such methods in Figure~\ref{fig:mmd_pavlue}. The first (Figure~\ref{fig:mmd_pavlue}a) uses the MMD approach taken by Kore et al.\cite{kore_empirical_2024} and implemented in Alibi-Detect. The second (Figure~\ref{fig:mmd_pavlue}b) uses $p$-values from $\chi^2$ tests (for discrete variables) and Kolmogorov-Smirnov tests (for continuous variables), the default metrics used by Evidently AI, combined using a Bonferroni multiple hypothesis correction.

Monitoring $p$-values without considering the associated effect size, especially when the number of samples can be large, is sensitive to distribution shifts that are likely inconsequential in practice, but statistically significant.
In real-world data, this results in plots that appear volatile and difficult to interpret.
By contrast, we find that distribution similarity methods accompanied by appropriate uncertainty bounds give values that better lend themselves to intuitive visualization of data drift. Unlike volatile $p$-values, methods that directly quantify distributional similarity offer more stable measures for drift quantification (see the \nameref{sec:mmc} section).

\subsection{Monitoring at Separate Points of Care}
\begin{figure}[ht]
    \centering
    \includegraphics[width=0.95\textwidth]{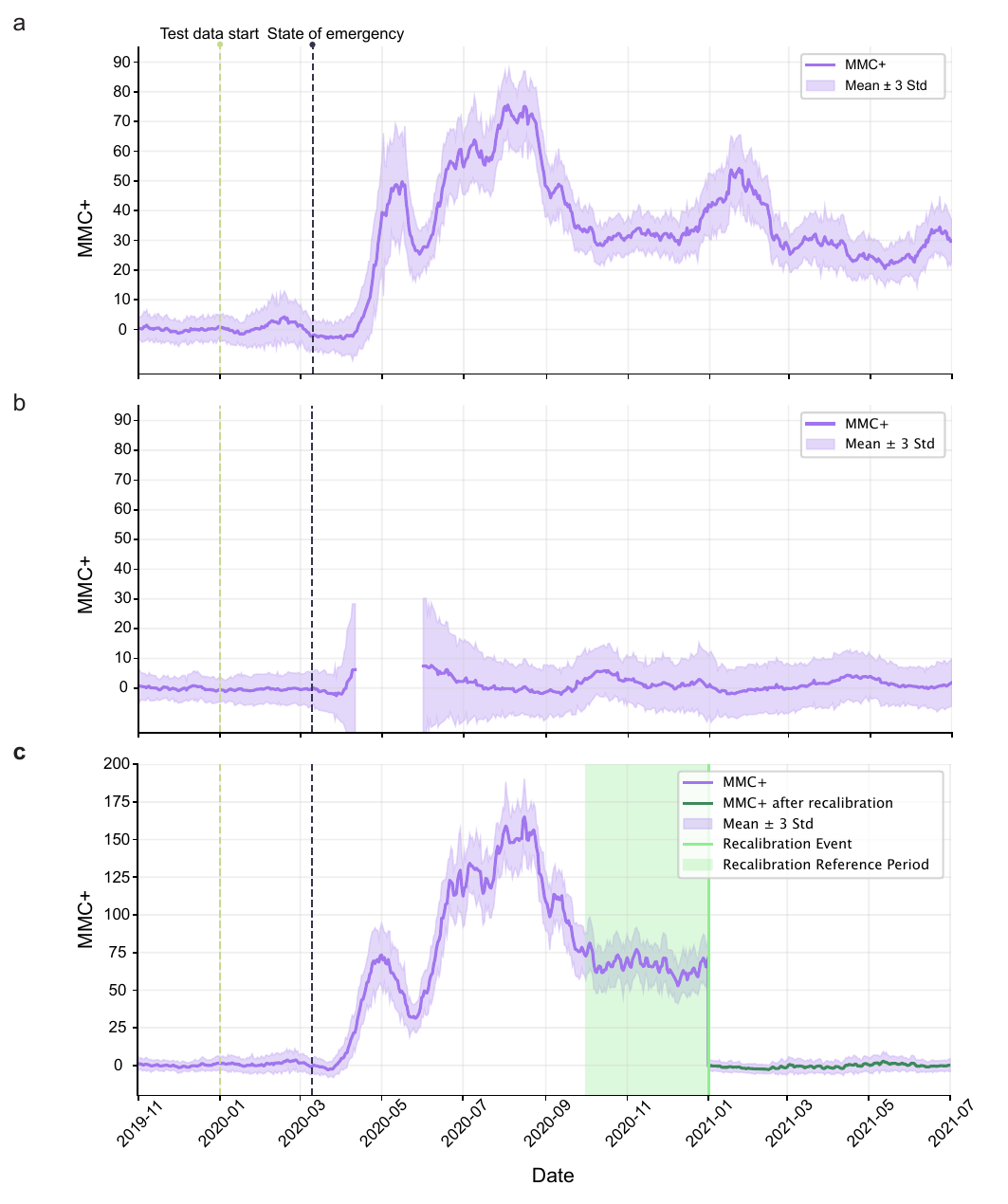}
    \caption{MMC+ at different points of care, and re-calibration. \textbf{a, b}: Plots that show the evolution of MMC+ over time for two different points of care: (a) the ER Point of Care and (b) the ambulatory center operating room Point of Care.  In both, the purple line represents the weighted MMC+ values, while the purple shaded area indicates the range or variability around the MMC+. Per-site monitoring surfaces drift patterns that an aggregate MMC+ could mask. At the ER (a), the pandemic shifts the patient population the model sees and MMC+ rises and stays elevated; at the ambulatory center (b), elective imaging drops off and the signal is a widening uncertainty band followed by a 50-day gap rather than a content shift. \textbf{c}: Plot of MMC+ over time with a re-calibration event. MMC+ recalibration on January 1st, 2021, using a new reference window from October 1st, 2020, to December 31st, 2020. The purple line shows MMC+ values before recalibration, and the dark green line indicates MMC+ values after recalibration. The purple shaded area represents the MMC+ range or variability. The recalibration event is marked by the vertical line, with the recalibration reference period highlighted in green. Because MMC+ is defined relative to a reference set, once a drift has stabilized and been audited, swapping in a new validated reference adapts the monitor to the new normal or to seasonal shifts.}
    \label{fig:mmc_other_poc}
\end{figure}

Hospitals typically use a wide range of X-ray devices at different points of care, such as operating rooms, emergency departments, and outpatient clinics. Each of these points of care may experience data drift events differently, as they often deal with distinct patient populations, imaging protocols, and workflows. Monitoring drift separately at each point of care is therefore crucial to gaining a complete and nuanced understanding of drift events. Figure~\ref{fig:mmc_other_poc} illustrates the behavior of the MMC+ metric for two different points of care within the Massachusetts General Hospital system.

Figure~\ref{fig:mmc_other_poc} a) shows the MMC+ metric for the emergency room, where a pronounced drift event occurs around the onset of the COVID-19 pandemic. This drift mirrors the broader trend observed in the overall MMC+, with the onset of the pandemic causing a sharp change in the datastream. By contrast, Figure~\ref{fig:mmc_other_poc} b) shows the MMC+ metric for an operating room in the ambulatory care center. Here, the drift pattern is noticeably different. Although the data changes around the same time, the drift is marked by a broadening of the uncertainty bounds indicating a reduction in the number of studies performed. Additionally, a 50-day gap in data collection follows, after which the data resumes without a sustained drift. This also showcases the importance of uncertainty bounds.

\subsection{Re-calibration After Drift}

One of the operational challenges in deploying AI models in clinical settings is the inevitability of data drift. Over time, the input data characteristics will shift, leading to changes in model performance. Eventually, these shifts stabilize, creating a new ``normal'' state. In such scenarios, consider a situation where data at a specific site has drifted, prompting an audit. Following this audit, necessary IT operations or model modifications are implemented to ensure that the model's performance returns to an acceptable range. However, to ensure continued monitoring effectiveness in this new data environment, it becomes essential to re-calibrate the MMC+ metric to track changes from this new reference set. Re-calibration of the MMC+ metric involves selecting a new reference set and recalculating the relevant weights and normalization factors (see the \nameref{sec:mmc} section).

This re-calibration is achieved by creating a new reference set using data from the audited period and recalculating the relevant weights and normalization factors. Our analysis as shown in Figure~\ref{fig:mmc_other_poc}  c) demonstrates that this approach is effective, as it successfully adjusts the MMC+ metric back into its acceptable range, thereby restoring model stability and ensuring consistent performance in the newly stabilized data environment.

\subsection{Effect of Detection Window Length}

\begin{figure}
    \centering
    \includegraphics[width=0.95\linewidth]{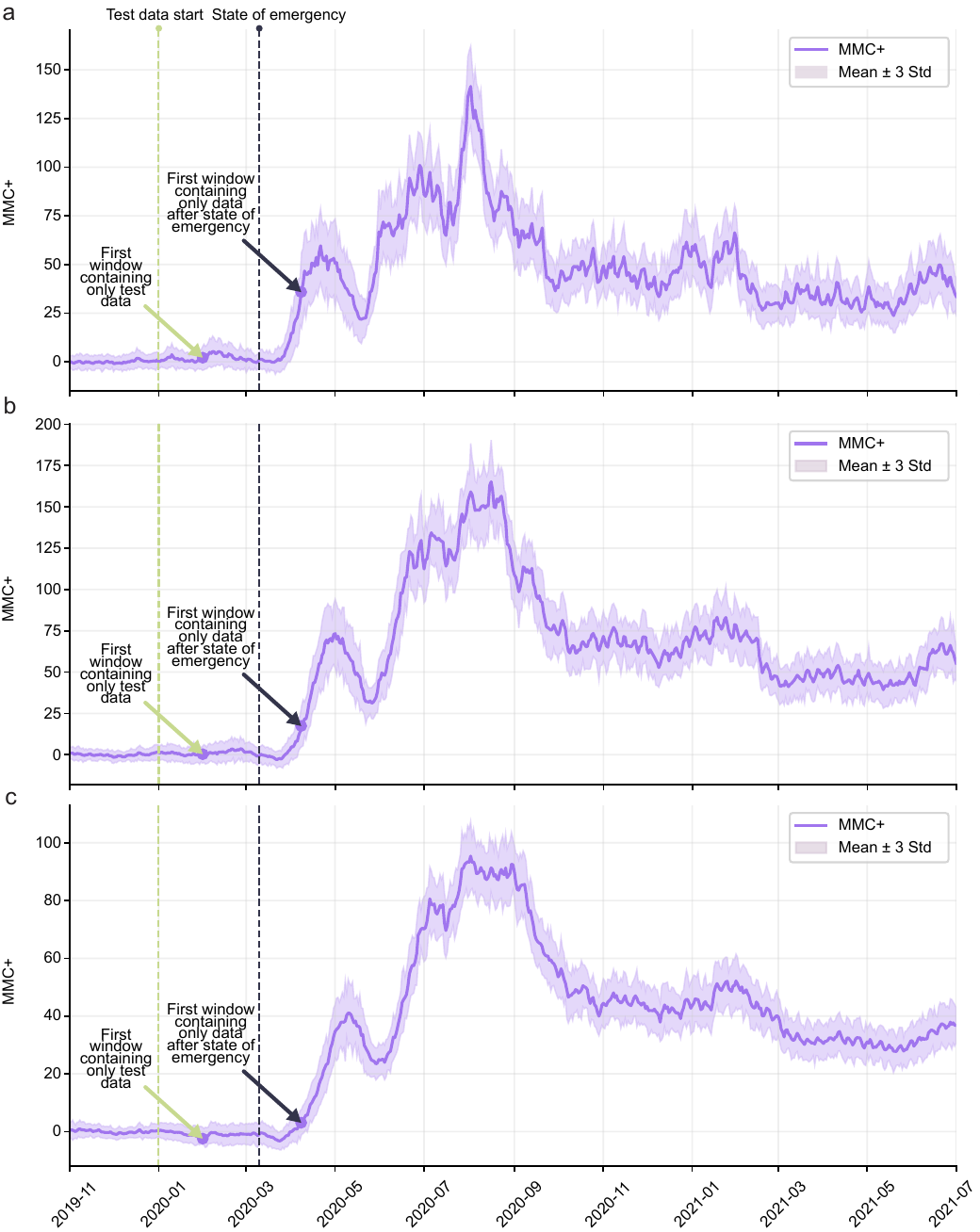}
    \caption{MMC+ calculated using different detection windows of different lengths. (a) 15-day, (b) 30-day, and (c) 45-day detection windows}
    \label{fig:mmc_window_size}
\end{figure}

In Figure~\ref{fig:mmc_window_size}, we repeat the experiment from Figure~\ref{fig:mmc} using window sizes of 15 days and 45 days, compared to 30 days in Figure~\ref{fig:mmc}. We observe a similar trade-off seen with all drift detection methods: increasing window time increases the lag time until drift events are detected and smooths out rapidly-occurring changes, but also reduces the uncertainty in the value of the drift metric. However, sites will also need to adapt the window size depending on the rate at which scans are acquired at that site. Sites with fewer scans may need to increase the detection window in order to reduce uncertainty in the value of MMC+.

\FloatBarrier
\section{Discussion}
Our work addresses the pressing need for continuous monitoring in medical imaging AI applications. The rapid integration of AI into imaging demands monitoring methods that extend beyond initial validation stages. Ongoing, interpretable monitoring is essential to promptly detect issues such as model drift, feature drift, and shifts in input data. Current practices, which often rely on infrequent reviews and user feedback, are insufficient. They are too slow to respond to emerging problems and lack the depth required to understand the nuances of model behavior. Real-time performance monitoring, while ideal, is frequently cost-prohibitive and may not effectively identify the underlying causes of performance changes.

 The need for adaptive monitoring was particularly evident during the COVID-19 pandemic, which brought rapid changes to healthcare workflows and emphasized the necessity of robust monitoring systems. As clinical protocols evolved and healthcare staff faced increased workloads, automated monitoring became crucial in identifying subtle shifts and ensuring the continued effectiveness of medical imaging AI. 
 However, beyond COVID-19, serial chest radiographs are common in hospitalized patients, and often have vast variations in acquisition techniques (projection - posterior-anterior versus anterior-posterior; portable versus non-portable acquisition devices and positions) and patient condition and abnormalities (variations in lung volumes, upright versus supine, rotation, support devices and catheters, and extent and type of abnormalities). These variations make interpretation of radiographs subjective and error-prone for radiologists and pose challenges for any AI tools.
 By providing real-time insights and easing the burden on medical professionals and administrators, these systems will play a vital role in navigating workload challenges while maintaining quality care.

Conventional drift monitors for medical imaging typically depend on encoders trained from scratch on local data, such as VAEs. This introduces two coupled problems. First, a locally trained encoder is fit to a narrow distribution and itself drifts as inputs evolve, contaminating the signal the monitor is meant to track. Second, valid temporal comparison requires a fixed embedding function: retraining the encoder invalidates the prior reference window and every distance computed against it. Together, these properties make per-site encoder training operationally costly and analytically fragile.

MMC+ addresses this by design, pairing modular per-component distribution comparison with a fixed, pre-trained foundation-model encoder \cite{codella2024medimageinsight} that produces a broad, stable description of medical images without per-site retraining. This embedding is deliberately a holistic, task-agnostic description of image appearance, used to flag any change in the distribution of images rather than to localize specific findings, while sensitivity to task-relevant change is carried by the model-output component. A consequence is that appearance drift is not attributed to specific pathological features, a deliberate trade-off for a stable, transferable representation. The same encoder transfers to a new site and an existing site's reference window remains valid as long as the encoder is held fixed. This is what makes the framework operationally scalable rather than a one-off configuration.

Our framework emphasizes essential properties that effective drift monitoring components should possess. Distribution comparison methods should provide stable quantification that decouples effect size from sample size, avoiding over-sensitivity to statistically significant but practically irrelevant shifts. Components should be robust across varying sample sizes and handle practical challenges such as evolving data characteristics, making them reliable for real-world clinical environments where protocols and populations change over time.

As previous works have noted\cite{kore_empirical_2024}, our findings demonstrate that data drift does not invariably correlate directly to performance degradation, and in some cases performance may even improve. This underscores the importance of monitoring both the model outputs and the input data to gain a comprehensive understanding of the system's behavior. We observed a clear connection between MMC+ and model performance, with stable performance while MMC+ remains within acceptable limits and variable, unpredictable performance that warrants audits when MMC+ deviates substantially. We also observe that monitoring different operational sites (such as ER vs PostOp) separately, when feasible, can enhance detection accuracy and provide more granular insights into the model's performance across diverse datasets. This insight is crucial as it offers a cost-effective compromise between continuous performance monitoring and periodic assessments.

In practical terms, hospitals can deploy MMC+ for continuous, real-time monitoring of AI models. This assists clinicians in identifying when AI support may be unreliable or require attention, thereby enhancing patient safety. However, obstacles remain. Existing methods might miss subtle yet important declines in model performance, and relying solely on proxies without verifying actual performance could compromise patient outcomes.

Beyond the architectural improvements, MMC+'s wider value is operational. Clinical AI deployments assume stationarity, that the data tomorrow looks like the data the model was trained on, but real clinical environments are not stationary. Protocols change, patient populations shift, scanners are replaced, and reporting practices evolve. MMC+ provides a way to quantify how far the current data has moved from the reference state, directly on the inputs the classifier sees and the outputs it produces. What counts as actionable drift, and the operational response it warrants, ultimately rests with each site's operators and IT policy. MMC+ supplies the measurement together with a mathematically grounded starting threshold, shown to match where the data would be split in retrospect, giving a defensible decision point. Indirect proxies, such as monitoring radiology report content, cannot separate true data drift from human-layer changes like new radiologists, reporting templates, or institutional policies. By measuring drift on the actual data and model behavior, MMC+ isolates the signal that matters for AI reliability from the noise of human and process change. Combined with the uncertainty bounds that handle fluctuations in study volume, this is what moves the framework from a research concept toward something that can run in clinical operations.

Our results highlight the importance of choosing drift quantification methods that separate statistical significance from practical relevance, particularly when large sample sizes can amplify inconsequential shifts. Methods that conflate these concerns often lead to volatile and difficult-to-interpret signals. In contrast, approaches that directly quantify distributional dissimilarity offer more stable and intuitive representations by measuring meaningful changes rather than significance levels. These findings suggest that measuring effect size rather than testing hypotheses presents a more effective foundation for tracking and visualizing drift in real-world applications.

While the MMC+ framework is designed for extensibility, with a modular architecture, a fixed foundation-model encoder, and per-component distribution comparison that accommodates new data types, we acknowledge that the empirical findings are limited to chest radiographs at a single institution and to a drift event dominated by the COVID-19 pandemic. This scope reflects the difficulty in finding clinical sites with real, labeled drift events. These events are rare and seldom shared publicly, and synthetic drift constructions tend to be contrived, rarely capturing the irregular, multi-source shifts seen in clinical practice. The COVID-19 period remains one of the few large-scale, well-documented, naturally occurring drift events available for study.

The framework itself imposes no assumptions or restrictions specific to chest radiography. Other two-dimensional modalities such as mammography can be operationally accommodated by substituting the reference data and recalibrating component weights. Extension to 3D formats like CT or MRI can be accommodated by replacing or augmenting the image embedding step; MedImageInsight has already been adapted for this purpose in 3D Medical Image Retrieval \cite{benabacha2026benchmarks}. The specific component weights, MMC+ threshold, and reference baseline reported here are properties of this dataset and site, and adapting MMC+ to a new context constitutes per-deployment recalibration rather than a generalization claim about the framework itself. The weighting uses a correlation heuristic to set the individual metadata feature weights, while the three top-level components are weighted equally, a simplification that emphasizes metadata features co-varying with performance on the reference window and not intended to model the exact form of that relationship.

MMC+ provides a scalable framework architecture for drift monitoring in medical imaging AI that accommodates appropriate methods for distribution comparison and leverages generalizable input representations. The framework addresses limitations of existing approaches by eliminating site-specific training requirements and enabling adaptation across various settings. Furthermore, the architectural design aligns with FDA guidance on continuous post-market surveillance for AI-enabled medical devices, providing signals when performance may deviate from acceptable bounds while meeting evolving regulatory standards. Future work should evaluate the framework across additional imaging modalities, characterize how component choices affect detection sensitivity, and address operational considerations such as alert thresholds and resource constraints. Overall, MMC+ advances dependable and scalable AI monitoring in healthcare by establishing principles for robust drift detection systems.

\section{Methods}
\label{sec:methods}

\subsection{Metric Development and Calculation}
\label{sec:mmc}

In our previous work, we introduced the CheXstray framework, which laid the foundation for generating the multi-modal concordance (MMC) metric.
In outline, MMC is an aggregate of distance measurements across multiple data elements including image metadata, image appearance, and model predictions, providing a single value representing the overall dissimilarity between two sets of medical exams.
For each component data element, a unimodal distance metric is calculated using appropriate statistical methods.
These distance metrics are then combined through a process of normalization, weighting, and aggregation to produce a unified MMC score.

In this work, we introduce an updated MMC+ framework that identifies essential properties for effective distribution comparison, incorporates uncertainty quantification to reflect sampling variability, and adopts foundation-model embeddings so that the framework can be applied across sites without per-site retraining of the image encoder. The framework is designed to be operationally extensible across modalities, sites, and tasks; the present study evaluates a single instantiation on a real-world clinical chest radiograph data stream. The following sections describe each component, the requirements it addresses, and the aggregation and normalization approach inherited from the original MMC.

The multi-modal concordance (MMC+) metric quantifies the degree of drift or change between a \emph{reference window}, $\refwindow$, representing a dataset with well-characterized model performance, and a \emph{detection window}, denoted by $\detwindow$, with unknown model behavior. 
The reference window is typically collected shortly before a model is deployed, using data with known ground truth to assess model performance. 
The detection window, on the other hand, consists of the set of samples within a sliding temporal window of fixed duration used to assess drift over time. 
If the MMC+ indicates a high level of dissimilarity between the detection window and the reference window, it suggests that the model is currently operating on data upon which its performance is not well characterized.

The detection window is defined by two key temporal parameters: duration and stride. The duration, $\duration$, specifies the total length of time that each window covers, while the stride, $\stride$, determines the time interval between the start of one window and the start of the next.
As such, if a data stream consists of items $x_i$ (each representing both an image and its associated metadata) arriving at time $t_i$, the detection window is constructed according to:
\begin{equation}
\begin{aligned}
   \detwindow_{k} &= \{x_i \mid t_k - \tau < t_i \leq t_k\} \\ 
   t_{k+1} &= t_k + \stride \\
\end{aligned}
\end{equation}

Each detection window \( \detwindow_{k} \) serves as the basis on which the MMC+ metric is computed. As the detection window slides over time, new windows are generated at each step, enabling continuous monitoring of model performance. The MMC+ metric quantifies the drift between these detection windows and the reference window, providing a robust measure of how well the model is performing on current data relative to the established baseline. By systematically calculating MMC+ for each detection window, we can assess and track the stability of the data stream over time.
For clarity, we drop the notation indicating the time at which the window was taken, as the metric calculation is generic with respect to time and does not depend on when the window was collected.

Distribution comparison is the core component for quantifying differences between the reference and detection windows.
The original MMC metric relied on statistical hypothesis tests for this purpose. However, further experimentation revealed limitations in this approach, prompting us to establish principles for effective drift quantification.

In practice, distribution comparison methods should satisfy several essential requirements. First, they should provide well-defined measures even when categorical values appear in one set but not the other, a common occurrence when new equipment or protocols are introduced. Second, they should quantify distributional dissimilarity in a way that is robust to sample size variations, separating the magnitude of shifts from statistical significance. Third, the methods should measure distributional differences directly rather than conflating distance with uncertainty estimation. For continuous variables, methods should handle multivariate data and detect shifts in joint distributions, capturing both changes in probability density and shifts in distribution location or shape. 

The MMC+ framework accommodates any comparison methods that satisfy these requirements. For this implementation, we employ non-parametric probability distribution comparison measures appropriate for both discrete categorical variables and continuous multivariate data, avoiding distributional assumptions while providing stable, interpretable quantification of distributional dissimilarity. Specific methods and their mathematical formulations are described in the \nameref{sec:mmc-components} section, with additional details in the Supplementary Information.

Effective drift monitoring requires quantifying not only distributional differences but also the uncertainty in those measurements, particularly when detection window sample sizes vary over time. The original CheXstray framework addressed sample size sensitivity through bootstrap oversampling to a fixed size, but this approach discards information about measurement uncertainty at low sample sizes.

MMC+ adopts an alternative approach that explicitly quantifies uncertainty in distribution comparison measures. Following the Flapjack algorithm\cite{papp2022bounds}, we repeatedly subsample from the reference window to estimate both the central tendency and variability of the comparison measure. Specifically, we draw two independent random subsets $U, V \subseteq \refwindow$ without replacement, where subset sizes match the detection window size: $\lvert U \rvert = \lvert V \rvert = \lvert W \rvert$. This ensures fair comparisons unbiased by differing sample sizes. The procedure requires $\lvert \detwindow \rvert < \lvert \refwindow \rvert$ for all detection windows, necessitating a sufficiently large reference window in practice. We repeat this sampling $N_s$ times to generate multiple subsets $\lbrace U_n \rbrace_{n=1}^{N_s}$ and $\lbrace V_n \rbrace_{n=1}^{N_s}$, which are used to compute a distribution of comparison measure estimates:
\begin{equation}
    \distset(\detwindow, \refwindow) = \lbrace \distsamp(\detwindow, \refwindow) \rbrace_{n=1}^{N_s} = \lbrace \mathcal{D}(U_n, \detwindow) - \mathcal{D}(U_n, V_n) \rbrace_{n = 1}^{N_s}
\end{equation}
where $\mathcal{D}$ denotes the appropriate comparison measure for each variable type. The baseline correction $\mathcal{D}(U_n, V_n)$ accounts for inherent variability within the reference window due to finite sampling, isolating the true distributional difference between $\detwindow$ and $\refwindow$ beyond what sampling variation alone would produce. The number of iterations $N_s$ can be adjusted to balance computational cost against precision. For notational simplicity, we subsequently denote this set as $\distset$, with the dependence on $\refwindow$ and $\detwindow$ implicit.

We report the mean value of $\distset$ over samples as the estimated distance and use the range of values within 3 standard deviations, calculated across $\distset$, of that value to obtain an estimate of the uncertainty of the distance metric. 
Since MMC+ aggregates comparison measures from multiple components, we apply this sampling procedure to each component independently, computing distance estimates using the appropriate measure for each data type. We denote component-specific measures with superscript $c$, where $\distisamp$ represents the distance estimate for the $c$\textsuperscript{th} component from the $n$\textsuperscript{th} subsample:
\begin{equation}
    \metricos^{(c)}  = \frac{1}{N_s}\sum_{n = 1}^{N_s} \distisamp, \quad
    \langle \metricos^{(c)}_{U}, \metricos^{(c)}_{L} \rangle = \langle \metricos^{(c)} + 3 \sqrt\frac{\sum{_{n = 1}^{N_s}(\distisamp-\metricos^{(c)})^2}}{n-1} , \metricos^{(c)} - 3 \sqrt\frac{\sum{_{n = 1}^{N_s}(\distisamp-\metricos^{(c)})^2}}{n-1} \rangle
\end{equation}
where $\metricos^{(c)}$ provides the central estimate of distributional dissimilarity for component $c$, while $\metricos^{(c)}_{U}$ and $\metricos^{(c)}_{L}$ capture the range of estimates across all $N_s$ subsamples, quantifying uncertainty in the measurement.

The MMC+ metric combines distance metrics from all components $c$ into a single metric. Since the metrics for each component have different ranges, we first normalize them.
For the $c$-th component, we calculate normalization parameters by finding the mean $\sshift^{(c)}$ and standard deviation $\sscale^{(c)}$ of the comparison measure $\metricos^{(c)}$ between each possible detection window $W_k$ within the reference set $R$ and the remainder of the reference set $\hat{R}_k$, defined as:
\begin{equation}
   \hat{R}_k = \{ x_i \mid x_i \in R,\ x_i \notin W_k \}
\end{equation}

In addition to normalization, each component is assigned a weight, and the weights sum to one. These weights can be set manually or derived from a heuristic that reflects a component's importance or reliability, such as the strength of its association with model performance.

For a weight derived from a component's association with performance, we compute a value \(\sweight^{(c)}\) from the absolute value of the Pearson correlation coefficient ($\rho$) between the distributional comparison measure and model performance. Here, performance is measured using the area under the receiver operating characteristic curve (AUROC), denoted as $a(W_k)$, where $W_k$ represents each detection window within the reference set $R$. Like the normalization calculation, the comparison measure is computed between $W_k$ and the remainder of the reference set, $\hat{R}_k$. This weight calibration is the only step in the framework that consumes ground-truth labels, and it does so on the reference window where labels are already required to characterize baseline performance; once \(\sweight^{(c)}\) is fixed, no further label access is required during monitoring. The absolute Pearson correlation is a simple, interpretable rule that measures how strongly a feature's drift co-varies with performance on the reference window, giving more weight to features that track performance. Because the weight captures only the strength of this association, not the form of the drift-performance relationship, a feature with a weak or non-linear association is down-weighted rather than mis-weighted. If monotone non-linear associations are a concern, a rank-based correlation can be substituted. The weight is computed as:
\begin{equation}
    \sweight^{(c)} = \left\lvert \rho\left( \left\{ \metricos^{(c)}(\hat{R}_k, W_k) \right\}_k, \{a(W_k)\}_k \right) \right\rvert
\end{equation}
The weights are normalized to sum to one before aggregation. In the instantiation evaluated here, the three top-level components, image appearance, model output, and metadata, are weighted equally, and the correlation heuristic above is used to distribute the metadata component's weight across its individual metadata features.

Combining the normalized and weighted measures from $C$ components, we calculate the multimodal concordance metric (MMC+) for the detection window $\detwindow$ as:
\begin{equation}
    \mmc+  =  \sum_{c = 1}^{C} \sweight^{(c)}\frac{\metricos^{(c)} - \sshift^{(c)}}{\sscale^{(c)}}
\end{equation}
To account for uncertainty, we also compute the upper and lower bounds of MMC+ denoted as
\begin{equation}
   \left\langle \mmc+_{U},\ \mmc+_{L} \right\rangle = \sum_{c=1}^{C} \sweight^{(c)}\, \frac{\left\langle \metricos^{(c)}_{U},\ \metricos^{(c)}_{L} \right\rangle - \sshift^{(c)}}{\sscale^{(c)}}
\end{equation}

Here, $\mmc+_{U}$ and $\mmc+_{L}$ are the upper and lower bounds of the distance metric for component $c$, obtained from the uncertainty estimates as described in the previous section. The three top-level components are weighted equally, and within the metadata component, features that co-vary with AUROC on the reference data receive greater weight, while the modular structure of the aggregate metric is preserved. The upper and lower bounds provide a range for MMC+, offering insight into potential variability due to sampling or measurement uncertainties.

The reference window's statistics define a ``no-drift'' baseline, so by construction it contains no drift-positive examples with which to calibrate an operating threshold empirically. We therefore set the operating threshold prospectively, using only the distribution of MMC+ within the reference set itself.

MMC+ is a drift metric based on a normalized aggregate distance measure. Normalization is constructed self-referentially by calculating leave-one-out distances among the windows reference within the reference set. This means that MMC+ is centered at zero statistically, and that MMC+ values within the unit range represent distances similar to a window you would see in the reference set. Under drift conditions, distance metrics will become larger and MMC+ will tend positive.

Since this zero-centered range is not itself an operational threshold, we still need a separate heuristic for how far MMC+ can rise before we call it drift. We use Cantelli's inequality for this, a one-sided bound that holds for any random variable with finite variance regardless of its distribution\cite{cantelli1928}. For threshold selection, we use the conservative unit-variance scale implied by the normalized weighted sum. For a target per-window false alarm probability $p$, Cantelli's inequality gives:
\begin{equation}
    T_p = \sqrt{\frac{1-p}{p}}.
\end{equation}
Using $p=0.01$ gives $T_p=\sqrt{99}\approx 9.95$, which for simplicity of interpretation we round to the operating threshold of 10 used in the experiments. This threshold is one principled operating point; deployments may choose a different $p$ to reflect local audit capacity and risk tolerance.

\subsection{Dataset}
\label{sec:dataset}

Our dataset was created by querying an institutional radiology database at Massachusetts General Hospital for all imaging studies matching the following criteria: 1) modality of ``XR'' (X-ray), 2) an institution-specific examination code with one of two values corresponding to ``X-Ray Chest 1 View'' and ``X-Ray Chest PA And Lateral 2 Views'', 3) a study date from within the two-year period between 2019-07-01 and 2021-07-01 (``yyyy-mm-dd'' format, inclusive), and 4) patient age of 18 years or older.
This dataset was chosen to reflect the full set of studies that would likely be routed to an X-ray analysis model if it were deployed in clinical practice at Massachusetts General Hospital, as well as to capture the period of time before, during, and after the initial wave of the COVID-19 pandemic.

This process yielded 90,713 studies of which 90,581 (from 60,062 distinct patients) were successfully retrieved from the institutional Picture Archive and Communication System (PACS) along with their radiology reports. We discarded lateral images, and used all remaining images within a study (some studies contained multiple images).
To create ground truth labels for this dataset, we used a large language model (QWEN 1.5-72B) to extract labels for each of a set of nine findings by few-shot prompting the model with a small number of positive and negative examples.
We characterized the accuracy of this process on a subset of this same dataset in previously published work\cite{dorfner_comparing_2024}, and found it to be highly accurate for most findings.
Following the CheXpert convention\cite{Irvin_2019_chexpert}, the nine selected findings were: atelectasis, cardiomegaly, edema, lung opacity, pleural effusion, pleural other, pneumonia, pneumothorax, and support devices.
We excluded the following findings due to inconsistency when extracting ground truth labels: consolidation, enlarged cardiomediastinum, fracture, lung lesion, and ``no finding''.
Within the MMC+ framework, ground-truth labels are required only to compute the metadata feature weights on the reference window (see the \nameref{sec:mmc} section); reference-window labels are typically available from the pre-deployment validation period. Once these weights are fixed, monitoring of new detection windows proceeds without additional label access.

We divided the dataset into three subsets based on the date of acquisition as follows: studies acquired in the first three months (July 2019 through September 2019) were used for model training and validation, studies from the next three months (October 2019 through December 2019) were used as reference set for the drift metric, and the remaining studies covering an 18-month period from the start of January 2020 until the end of June 2021 were used for the drift experiments.
This division was chosen to approximately reflect a model development process. In such a process  a model would first be developed on retrospective images, then validated on recent images before being deployed prospectively on new images. Characteristics of the three dataset partitions are presented in Table \ref{tab:dataset}, and the distribution of findings across partitions is presented in Table~S2 in the Supplementary Information.

\begin{table}[ht]
    \centering
    \begin{tabular}{lccc}
         & \textbf{Training} & \textbf{Reference}\textsuperscript{*} & \textbf{Test}\\ \toprule
        Start Date & 2019-07-01 & 2019-10-01 & 2020-01-01\\
        End Date & 2019-09-30 & 2019-12-31 & 2021-07-01\\ 
         Total Images & 12,478 & 14,936 & 78,181\\ \midrule
        Age (years, mean $\pm$ std) & 60.2 $\pm$ 18.48 & 59.53 $\pm$ 18.22 & 58.64 $\pm$ 17.87 \\
        Female (\%) & 48.12 & 49.44 & 46.03 \\
        Inpatient (\%) & 14.74 & 14.44 & 27.84 \\
        Emergency (\%) & 31.53 & 29.37 & 27.68 \\
        Portable (\%) & 28.15 & 27.79 & 51.38 \\ \bottomrule
    \end{tabular}
    \caption{Summary of dataset characteristic statistics. \textsuperscript{*}The reference partition also serves as the model's validation set.}
    \label{tab:dataset}
\end{table}

\subsection{Classification Model}
\label{sec:classifier}

For the purposes of our experiments, the AI model whose performance we monitored was a classification model trained to detect multiple findings in a chest radiograph.
This model was based on the DenseNet-121 architecture, pretrained on the ImageNet-1k dataset \cite{huang2018densely, deng2009imagenet}.
The output layer was adapted to consist of nine elements, one for each of the nine findings, followed by a sigmoid activation to yield nine independent (multi-label) predictions.
The model was trained on the images from the training set, defined in the \nameref{sec:dataset} section. Training was conducted with a batch size of 64, a learning rate of $1 \times 10^{-4}$, a binary cross-entropy loss function, the Adam optimizer and a ReduceLROnPlateau learning rate schedule. The model was trained exclusively on frontal (AP and PA) images (lateral views were excluded) across all points of care. Training was performed using a single A100-40GB GPU (NVIDIA, Santa Clara, CA, USA).

\subsection{MMC+ Experimental Implementation}
\label{sec:mmc-components}

The preceding sections describe the architectural principles and requirements of the MMC+ framework. This section details how we instantiate the framework for our chest X-ray experiments, specifying the concrete methods and data elements used to satisfy those requirements.

For distribution comparison of continuous variables, we employ the Wasserstein distance\cite{panaretos2019statistical}, which effectively captures differences in multivariate distributions and handles shifts in both location and shape. For discrete categorical variables, we use the Hellinger distance\cite{oosterhoff2012note}, which provides well-defined measures even when categories differ between sets and remains stable across varying sample sizes. The full mathematical formulations of both metrics are provided in the Supplementary Information for reference.

MMC+ aggregates information from three types of data elements: image metadata, image appearance, and model predictions. While the framework is general and can accommodate various measurements, we describe here our specific instantiation for chest X-ray monitoring.

Every radiological image acquired in clinical practice is stored within a standardized DICOM (Digital Imaging and Communications in Medicine) file alongside various metadata attributes that describe acquisition parameters of the image, patient characteristics, and the clinical context of the acquisition.
For our experiments, we selected a number of DICOM attributes that may affect model performance on chest radiographs and used each as an individual element within MMC+.
A small number of additional attributes were sourced from the institutional Radiology Information System (RIS).
Table~S1 in the Supplementary Information lists the relevant attributes. Each metadata attribute is treated as either a discrete or continuous variable and compared using the appropriate distance metric (Hellinger or Wasserstein).

In our previous implementation, we used a 128-dimensional latent vector produced by a VAE to represent the appearance of medical images. The VAE was trained on chest radiographs from the PadChest dataset \cite{padchest}, and the resulting latent vector served as the basis for calculating distance metrics to monitor shifts in image data. While this approach provided a functional, low-dimensional representation, it posed several challenges. Training the VAE required significant computational resources and data preparation that, in operation, would need to be repeated at each site. Moreover, the VAE's limited expressiveness made it insufficient for capturing the complexity of medical images, particularly across imaging modalities.

To overcome these limitations, we transitioned to using MedImageInsight \cite{codella2024medimageinsight}, a foundation model designed specifically for medical image analysis, which provides a 1024-dimensional embedding. Unlike the VAE, MedImageInsight does not require site-specific training, offering a powerful pre-trained solution that can be applied to a wide range of medical images out of the box. The 1024-dimensional embedding is more expressive than the prior VAE latent and can describe a broad range of medical images out of the box, without per-site retraining of the encoder.

By leveraging MedImageInsight’s pre-trained embeddings, we remove the per-site encoder retraining that the prior VAE pipeline required. MedImageInsight is a foundation model purpose-built for medical-image description, with reported validation across CT, MRI, mammography, and text \cite{codella2024medimageinsight}, so the encoder transfers to a new site without modification. Adapting MMC+ as a whole to a new site still requires curating a reference window and recalibrating component weights; the framework is designed to support this re-instantiation rather than to be deployed unchanged. We compute the Wasserstein distance on these 1024-dimensional embeddings to quantify appearance drift.

The raw classifier output (after the sigmoid activations) consists of $9$ independent values in the range $0$--$1$, one for each of the findings.
These values carry information about both the likely class distribution of the datastream and the model's confidence in its predictions.
This nine element vector is used directly as a multidimensional variable, with the Wasserstein distance computed on this continuous multivariate representation to quantify drift in model predictions.

\FloatBarrier

\section*{Data Availability}

The datasets generated and/or analyzed during the current study are not publicly available due to laws and hospital policies related to patient privacy.

\section*{Code Availability}

Source code for the MMC+ framework is publicly available at \url{https://github.com/FJDorfner/MedImaging-ModelDriftMonitoring/tree/MMC_plus}.

\section*{Acknowledgements}

No funding was received for this research. We acknowledge Alexandra Eikenbary for creating Figure 1.

\section*{Author Contributions}

J.M., C.P.B., M.P.L., I.T. and A.E. conceived the study. J.M., F.J.D., X.Y., and C.P.B. designed and conducted experimental studies. G.D. and M.K. prepared data and contributed to the study design. J.M., F.J.D., and C.P.B. wrote the main manuscript text. All authors reviewed and approved the manuscript.

\section*{Competing Interests}

J.M., A.E., M.P.L., and I.T. are employees of Microsoft, which develops AI monitoring solutions potentially related to the subject of this manuscript. I.T., J.M., Stephen Kaiser, Smitha Srinivasa Saligrama, Steven Joel Borg, M.P.L., and Arjun Soin are inventors on pending U.S. patent application US 18/051,786, ``Data Set Distance Model Validation,'' assigned to Microsoft Technology Licensing, LLC. The application concerns dataset-distance-based validation and monitoring of inferential models, including aspects of the drift-monitoring framework described in this manuscript. The other authors declare no competing financial or non-financial interests.

\bibliography{references}

\end{document}

\maketitle

\section*{Supplementary Tables}

\begin{table}[H]
    \centering
    \resizebox{\textwidth}{!}{%
    \begin{tabular}{l|l|c|cc|l}
        \textbf{Feature} & \textbf{Type} & \textbf{Weight} & \multicolumn{2}{c|}{\textbf{Normalization}} & \textbf{Description} \\
        & & & $\boldsymbol{\zeta}$ & $\boldsymbol{\eta}$ & \\ \toprule
        \multicolumn{6}{l}{\textbf{Metadata}} \\ \midrule
        \hspace{1em}Modality & Discrete & 0.0124 & 1.7e-3 & 1.2e-3 & Tag: \texttt{0008,0060}. DX ``Digital Radiography'' or CR ``Computed Radiography''. \\
        \hspace{1em}Manufacturer & Discrete & 0.0271 & 7.1e-3 & 4.3e-3 & Tag: \texttt{0008,0070}. Manufacturer of the scanner that acquired the image. \\
        \hspace{1em}ManufacturerModelName & Discrete & 0.0273 & 6.4e-3 & 3.8e-3 & Tag: \texttt{0008,1090}. Model name of the scanner that acquired the image. \\
        \hspace{1em}Patient Sex & Discrete & 0.0224 & 1.4e-3 & 1.1e-3 & Tag: \texttt{0010,0040}. Sex of the patient. \\
        \hspace{1em}KVP & Continuous & 0.0047 & 0.20 & 0.11 & Tag: \texttt{0018,0060}. Kilo-volt peak of the source X-Ray tube. \\
        \hspace{1em}XRayTubeCurrent & Continuous & 0.0134 & 2.4 & 1.3 & Tag: \texttt{0018,1151}. Current of the source X-Ray tube. \\
        \hspace{1em}ExposureIn$\mu$As & Continuous & 0.0053 & 43 & 24 & Tag: \texttt{0018,1153}. Radiation exposure in $\mu$As. \\
        \hspace{1em}ViewPosition & Discrete & 0.0008 & 7.4e-3 & 4.2e-3 & Tag: \texttt{0018,5100}. Radiographic view (e.g.\ posterior/anterior, lateral). \\
        \hspace{1em}PhotometricInterpretation & Discrete & 0.0434 & 5.4e-3 & 5.6e-3 & Tag: \texttt{0028,0004}. Interpretation of pixel data. \\
        \hspace{1em}Rows & Continuous & 0.0394 & 4.2 & 3.8 & Tag: \texttt{0028,0010}. Number of rows in the image. \\
        \hspace{1em}Columns & Continuous & 0.0252 & 3.3 & 2.5 & Tag: \texttt{0028,0011}. Number of columns in the image. \\
        \hspace{1em}PixelSpacing & Continuous & 0.0191 & 1.2e-4 & 7.4e-5 & Tag: \texttt{0028,0030}. Physical spacing between pixels. \\
        \hspace{1em}BitsStored & Discrete & 0.0458 & 6.0e-3 & 4.7e-3 & Tag: \texttt{0028,0101}. Pixel bit-depth. \\
        \hspace{1em}Patient Age & Continuous & 0.0015 & 0.15 & 0.092 & Age of the patient in years. \\
        \hspace{1em}Point of Care & Discrete & 0.0105 & 2.9e-3 & 1.8e-3 & Hospital location. \\
        \hspace{1em}Is Stat & Discrete & 0.0314 & 4.5e-3 & 4.9e-3 & Whether image was marked for urgent review. \\
        \hspace{1em}Exam Code & Discrete & 0.0037 & 1.3e-3 & 6.9e-4 & Internal procedure code (see the Dataset section of the main manuscript). \\ \midrule
        \textbf{Image Appearance} & Continuous & $\sfrac{1}{3}$ & 0.088 & 2.3e-3 & Foundation-model image-appearance embedding. \\
        \textbf{Model Output} & Continuous & $\sfrac{1}{3}$ & 6.5e-3 & 7.8e-4 & Model output class probabilities. \\
        \bottomrule
    \end{tabular}%
    }
    \caption{Monitored components and per-feature metadata attributes used for experiments, listing the keyword, type (discrete or continuous), the MMC+ weight, the reference normalization parameters $\zeta$ and $\eta$, and a description. Normalization values are rounded to two significant digits; scientific notation is used where it is no longer than decimal notation. The three top-level components (metadata, image appearance, and model output) are weighted equally. The metadata weights shown are distributed across the individual metadata features by the correlation heuristic and sum to $\sfrac{1}{3}$.}
    \label{tab:dicom_metadata}
\end{table}

\begin{table}[ht]
    \centering
    \begin{tabular}{lccc}
         & \textbf{Training} & \textbf{Reference} & \textbf{Test}\\ \toprule
        Total Images & 12,476 & 14,931 & 78,168\\ \midrule
        Atelectasis & 3,610 (28.94\%) & 4,204 (28.16\%) & 25,966 (33.22\%) \\
        Cardiomegaly & 1,838 (14.73\%) & 2,048 (13.72\%) & 10,381 (13.28\%) \\
        Edema & 1,455 (11.66\%) & 1,582 (10.60\%) & 12,022 (15.38\%) \\
        Lung Opacity & 5,554 (44.52\%) & 6,479 (43.39\%) & 38,252 (48.94\%) \\
        Pleural Other & 637 (5.11\%) & 717 (4.80\%) & 3,810 (4.87\%) \\
        Pleural Effusion & 2,239 (17.95\%) & 2,491 (16.68\%) & 16,742 (21.42\%) \\
        Pneumonia & 1,668 (13.37\%) & 2,061 (13.80\%) & 13,194 (16.88\%) \\
        Pneumothorax & 287 (2.30\%) & 350 (2.34\%) & 2,548 (3.26\%) \\
        Support Devices & 3,190 (25.57\%) & 3,645 (24.41\%) & 25,694 (32.87\%) \\ \bottomrule
    \end{tabular}
    \caption{Distribution of findings across training, reference, and test datasets. Values represent the number of positive cases with prevalence percentages in parentheses.}
    \label{tab:label_distribution}
\end{table}

\section*{Supplementary Note: Distance Metric Formulations}

This section provides the mathematical formulations for the distance metrics used in our implementation of MMC+.

For discrete variables (such as scanner model name or photometric interpretation) with $K$ discrete values, we calculate distances between two generic sets of data points, $P = \lbrace p_i \rbrace_{i=1}^{N}$ and $Q = \lbrace q_i \rbrace_{i=1}^{M}$ where $p_i, q_i \in \lbrace 1, 2, \ldots, K \rbrace$, using the Hellinger distance\cite{oosterhoff2012note}, which is defined as
\begin{equation}
    \mathcal{H}(P, Q) = \sqrt{\frac{1}{2}\sum_{k=1}^K \left( \sqrt{\frac{1}{N}\sum_{i=1}^N \mathbb{I}\left(p_i = k\right)} - \sqrt{\frac{1}{M}\sum_{i=1}^M \mathbb{I}\left(q_i = k\right)} \right)^2},
\end{equation}
where $\mathbb{I}(\cdot) \in \lbrace 0, 1 \rbrace$ is the indicator function.

For $d$-dimensional continuous variables, $p_i, q_j \in \mathbb{R}^d$, we use the Wasserstein distance (otherwise known as the earth-mover's distance).
In its general form, the Wasserstein distance between two distributions $p(x)$ and $q(x)$ is given by the cost of the optimal transport plan that moves the probability mass of $p(x)$ to $q(x)$, where the cost is defined as equal to the amount of probability mass moved multiplied by distance moved in the underlying space\cite{panaretos2019statistical}.
For empirical distributions, where the distribution is formed of point masses at sample points $P = \lbrace p_i \rbrace_{i=1}^{N}$ and $Q = \lbrace q_j \rbrace_{j=1}^{M}$, the optimal transport plan, and therefore Wasserstein distance, may be found by computing Euclidean distances between all pairings of points in $P$ and $Q$, and solving the following linear program:
\begin{equation}
   \begin{aligned}
        \mathcal{W}(P, Q) \quad& = & \min_{\gamma} & \sum_{i=1}^N \sum_{j=1}^M \gamma_{i,j} \lVert p_i - q_j \rVert_2  \\
        &&\textrm{s.t.} \quad & \sum_{i=1}^N \gamma_{i,j} = \frac{1}{M}, \quad  j = 1, \dots, M \\
        &&& \sum_{j=1}^M \gamma_{i,j} = \frac{1}{N}, \quad  i = 1, \dots, N \\
        &&&0 \leq \gamma_{i,j} \leq 1, \quad  i = 1, \ldots, N, j = 1, \ldots, M \\
    \end{aligned}
\end{equation}
where $\gamma_{i,j} \in \mathbb{R}$ is the quantity of probability mass moved from initial point $p_i$ in the first distribution to end point $q_j$ in the second distribution.
In our implementation, we use the open-source Python Optimal Transport (POT) library \cite{flamary2021pot} to calculate the solution to this optimal transport problem.

\bibliographystyle{unsrt}
\bibliography{references}